\documentclass[fleqn,10pt,onecolumn]{wlscirep}

\usepackage[utf8]{inputenc}
\usepackage[T1]{fontenc}
\usepackage{multirow}
\usepackage{dutchcal}
\usepackage{svg}
\usepackage{threeparttable}
\usepackage{marvosym} 
\usepackage{lineno} 
\usepackage{eucal} 
\usepackage{tikz}
\usetikzlibrary{arrows.meta,positioning,fit,calc}
\usepackage{graphicx}
\graphicspath{{../images/}{images/}{./}}
\title{An adaptive and evolvable deep reinforcement learning framework for weather prediction}
\author[1]{Qiang Wu}
\author[1]{Han Li}
\author[1\Letter]{Jianping Huang}

\affil[1]{Collaborative Innovation Center for Western Ecological Safety, Lanzhou University, Lanzhou 730000, China}

\affil[\Letter]{email: hjp@lzu.edu.cn}

\begin{abstract}
No single AI weather model excels at all variables, pressure levels, and lead times.
Rather than building yet another architecture, we reframe the forecasting problem as one of coordination.
Here we present Feitian Adaptive Ensemble Weather (FTAE-Weather), a lightweight framework that learns, through deep reinforcement learning, when and where to trust each member of an open pool of pretrained forecasters.
A tactical Weight-Agent reads the current atmospheric state and assigns variable- and horizon-specific fusion weights, while a strategic Evolve-Agent periodically prunes underperforming models and absorbs newly released ones.
Asynchronous prediction caching keeps training cost independent of the slowest constituent model.
Adding fewer than 0.01\% extra parameters, FTAE-Weather reduces RMSE by from 17.2\% to 78.3\% over the best individual model in 10 atmospheric variables and outperforms conventional ensemble baselines across lead times from 72 to 360~hours.
The framework thus converts a growing, fragmented inventory of specialist models into a single prediction system that strengthens as the field of AI weather forecasting releases new architectures---turning model diversity from a coordination challenge into a compounding scientific advantage.
\end{abstract}
\begin{document}

\flushbottom
\maketitle
\thispagestyle{empty}
\section*{Introduction}

Weather forecasts save lives and shape trillion-dollar decisions in agriculture, energy, and transportation\cite{endtoend2025weather,bauer2015quiet}.
For decades, higher-resolution numerical weather prediction (NWP) was the dominant path to better forecasts\cite{ritchie1995semiLagrangian, bauer2015quiet}.
That path is now reaching diminishing returns, as computational cost scales steeply with grid spacing while sub-grid processes remain difficult to represent\cite{bauer2020scalability,buizza2005comparison}.

A parallel revolution in machine learning has upended this trajectory\cite{lecun2015deep}.
Trained on petabytes of ERA5 reanalysis\cite{hersbach2020era5}, data-driven models now rival or surpass NWP on standard metrics\cite{pathak2022fourcastnet_pasc,bi2023panguweather,lam2023graphcast,chen2023fengwu,chen2023fuxi,kochkov2024neuralgcm,price2025gencast,bonev2025fourcastnet3,nguyen2023climax}, yet they disagree profoundly on \emph{which} aspects of the atmosphere they predict best.
Transformers\cite{bi2023panguweather, bodnar2024aurora, chen2023fengwu,vaswani2017attention} capture long-range teleconnections; graph neural networks\cite{lam2023graphcast,lang2024aifs,keisler2022forecasting} respect spherical geometry; diffusion models\cite{price2025gencast,pathak2024stormcast, harris2025residual} quantify forecast uncertainty; physics-informed designs\cite{kochkov2024neuralgcm,verma2024climode} enforce conservation laws; neural operators\cite{pathak2022fourcastnet_pasc,bonev2025fourcastnet3,li2021fourier} learn solution maps of partial differential equations.

The result is a paradox of plenty.
Benchmarks\cite{rasp2024weatherbench2,ben2024rise,charlton2024ai} reveal that Pangu-Weather\cite{bi2023panguweather} leads for upper-air circulation and tropical cyclone tracking\cite{Bremnes2024, Olivetti2024}, GraphCast\cite{lam2023graphcast} for near-surface variables at short-to-medium ranges\cite{Radford2025}, and FourCastNet\cite{bonev2025fourcastnet3} for extended integrations, yet no single model wins everywhere\cite{hagedorn2005rationale,tebaldi2007multimodel,doblas2005multimodel}.
Simple averaging or static weighting\cite{gneiting2005weather,fan2022spatiotemporal,weigel2010risks} cannot exploit this complementarity because they apply the same combination rule regardless of the atmospheric regime\cite{troin2021generating}.
Scaling individual architectures further offers diminishing returns at rising cost\cite{bonavita2024limitations,willard2025analyzing,bodnar2024aurora,bonev2025fourcastnet3}, and mixture-of-experts approaches\cite{jacobs1991adaptive,yuksel2012twenty,shazeer2017outrageously,chakraborty2025moe} in weather forecasting have so far relied on fixed gating rather than state-conditioned weighting.

We reason that the missing ingredient is an agent that can read the atmosphere and decide, in real time, which model to trust for each variable, location, and lead time.
Deep reinforcement learning (DRL)\cite{mnih2015humanlevel,lillicrap2015continuous,silver2016mastering} provides a natural formalism for this problem. The agent observes the current atmospheric state together with competing forecasts, selects continuous combination weights as actions, and refines its policy from delayed verification signals\cite{silver2016mastering,haarnoja2018soft,fujimoto2018addressing}.

Here we introduce Feitian Adaptive Ensemble Weather (FTAE-Weather), a framework that reframes weather prediction as a coordination problem rather than a winner-take-all contest.
FTAE-Weather does not train yet another foundation model; it learns to orchestrate existing ones via DRL\cite{sutton2018reinforcement,Fu_Wu_AAAIRL_2022}.
A Weight-Agent reads the current atmospheric state and generates variable- and horizon-specific fusion weights; an asynchronous prediction cache decouples policy updates from the differing inference speeds of constituent models.
A slower-acting Evolve-Agent monitors the accumulated weight evidence across evaluation epochs and swaps underperforming models for newly released architectures, keeping the pool current without manual curation.

Across lead times from 24 to 360~hours, FTAE-Weather outperforms every constituent model and standard ensemble baselines while adding fewer than 0.01\% extra parameters.
This result reveals a broader implication for the field. The rapid proliferation of specialist AI weather models, often viewed as a coordination burden, becomes a compounding scientific asset when models are treated as cooperating contributors rather than competing alternatives.
Each new architecture released by the field, regardless of design philosophy or training institution, can be absorbed into an FTAE ensemble and strengthen the collective forecast at negligible additional cost.
The framework thus opens a path in which progress in AI weather forecasting is measured not only by individual model performance but by the collective capability that emerges when diverse architectures are systematically coordinated.

\section*{Feitian Adaptive Ensemble Weather framework}
\label{sec:model}

\begin{figure*}[h]
    \centering
    \includegraphics[width=\linewidth]{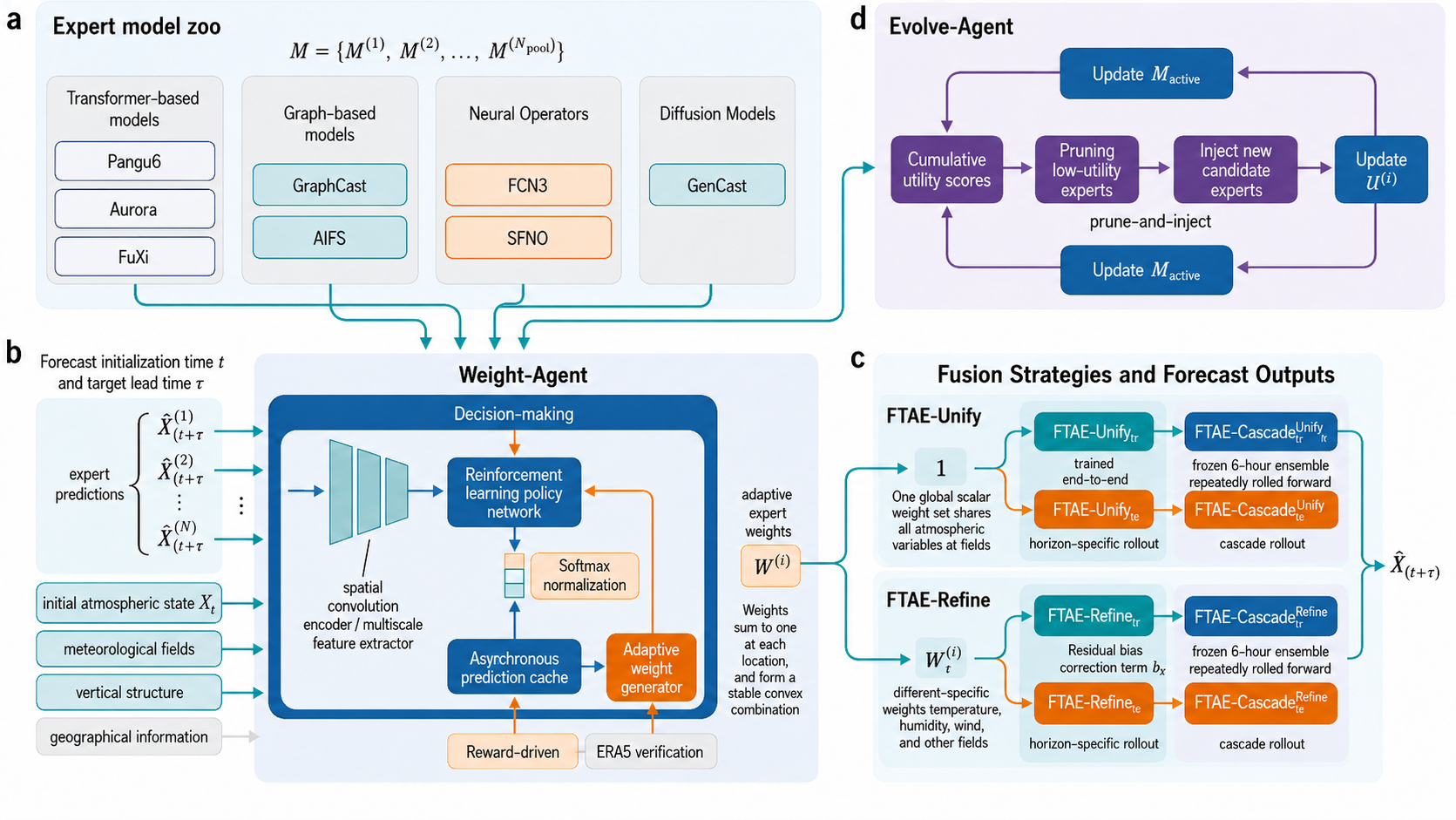}
\caption{
Overview of the FTAE-Weather framework.
\textbf{a},~The expert model zoo contains eight pretrained forecasters, including Transformer-based models (Pangu-Weather, Aurora, and FuXi), graph-based models (GraphCast and AIFS), neural-operator or spectral-operator models (FourCastNet3 and SFNO), and the diffusion-based probabilistic model GenCast.
\textbf{b},~The Weight-Agent receives the initial atmospheric state and the expert predictions, extracts multiscale spatial features, and generates adaptive expert weights through a reinforcement-learning policy network with Softmax normalization.
\textbf{c},~The expert forecasts are fused through two strategies: FTAE-Unify, which applies one shared set of scalar weights across all variables, and FTAE-Refine, which allows variable-dependent weights with an additional residual bias-correction term. Their cascade variants use the corresponding short-horizon ensemble recursively for longer lead times.
\textbf{d},~The Evolve-Agent updates the active model pool on a slower timescale by evaluating accumulated utility scores, pruning low-utility experts, injecting candidate experts, and updating the active set $\mathcal{M}_{\mathrm{active}}$.
}
\label{fig:framework}
\end{figure*}

The central idea behind FTAE-Weather is to treat forecast fusion not as a fixed recipe but as a sequential decision problem that a reinforcement learning agent solves anew for each atmospheric state (Fig.~\ref{fig:framework}).
The framework has three components: (i)~an open library of pretrained forecasting models (expert model zoo), (ii)~a tactical Weight-Agent that generates situation-dependent fusion weights via deep reinforcement learning\cite{mnih2015humanlevel,lillicrap2015continuous}, and (iii)~a strategic Evolve-Agent that curates the active model pool over time.

\subsection*{Expert model zoo}
The framework maintains a library of pretrained models $\mathcal{M}=\{M^{(1)}, M^{(2)}, \ldots, M^{(N_{\mathrm{pool}})}\}$ drawn from diverse architectural families: Transformer-based models (Pangu-Weather\cite{bi2023panguweather}, Aurora\cite{bodnar2024aurora}, FuXi\cite{chen2023fuxi}), graph-based models (GraphCast\cite{lam2023graphcast}, AIFS\cite{lang2024aifs}), diffusion-based probabilistic models (GenCast\cite{price2025gencast}), and neural-operator or spectral-operator models (FourCastNet3\cite{bonev2025fourcastnet3}, SFNO\cite{bonev2023sfno}).
At any time an active subset $\mathcal{M}_{\mathrm{active}} \subseteq \mathcal{M}$ of $N$ models participates in forecast generation.
Each model is treated as a black-box forecaster through a common input--output interface, so the eight pretrained models can be evaluated and combined without modifying or retraining their internal parameters.

\subsection*{Tactical generation via Weight-Agent}
The Weight-Agent produces fusion weights conditioned on the expert predictions, the atmospheric state, and the forecast lead time.
At initialisation time $t$ for target lead time $\tau$, each active model $M^{(i)} \in \mathcal{M}_{\mathrm{active}}$ produces a forecast $\hat{X}^{(i)}_{t+\tau}$ from the initial conditions $X_t$.
The Weight-Agent receives the concatenated expert predictions $\{\hat{X}^{(1)}_{t+\tau}, \ldots, \hat{X}^{(N)}_{t+\tau}\}$ together with $X_t$ and assigns a weight $W^{(i)}$ to each model.
Weight generation is framed as a reinforcement learning problem: the agent observes the expert forecasts and the ERA5 verification state, selects weight configurations as actions, receives a reward based on forecast skill, and updates its policy to maximise expected return.
A convolutional encoder captures multiscale atmospheric patterns and inter-model discrepancies; a Softmax layer enforces the simplex constraint so that weights sum to one at every grid point.
An asynchronous caching mechanism stores precomputed model outputs, decoupling policy updates from the differing inference speeds of constituent models.
FTAE-Weather supports two fusion granularities.

\paragraph{FTAE-Unify.}
All atmospheric variables share a single set of scalar weights $\{w^{(1)}_{\tau}, \ldots, w^{(N)}_{\tau}\}$ with $w^{(i)}_{\tau} \in [0,1]$ and $\sum_i w^{(i)}_{\tau} = 1$.
Temperature, humidity, and wind fields all receive the same expert combination.
Because a single set of scalar weights is shared across all atmospheric variables, it cannot represent variable-specific systematic offsets; introducing a per-variable bias term $b_{\tau}$ would be redundant and under-constrained in this setting.
Accordingly, the bias correction term is fixed at zero ($b_{\tau} = 0$), and the fused forecast reduces to a strict convex combination: $\hat{X}_{t_0+\tau} = \sum_{i=1}^{N} w^{(i)}_{\tau} M^{(i)}(X_{t_0})$.
The Weight-Agent learns these global weights by maximising mean forecast skill across all variables.

\paragraph{FTAE-Refine.}
Different atmospheric variables receive distinct weight allocations $W^{(i)}_{\tau}$, allowing the agent to exploit the fact that model skill varies across fields.
For example, temperature and specific humidity may be weighted towards different experts.
A residual correction term $b_{\tau}$ compensates for systematic biases remaining in the weighted combination.
The agent optimises variable-dependent weights by maximising per-field forecast skill metrics.

%

\paragraph{FTAE-Unify$_{\tau}$ and FTAE-Refine$_{\tau}$.}
For a target horizon $\tau$ (e.g.\ 24, 72, or 120~h), each expert model $M^{(i)}$ rolls out its own autoregressive chain ($X_t \rightarrow X_{t+6} \rightarrow \cdots \rightarrow X_{t+\tau}$) independently.
The Weight-Agent sees only the final outputs $\{\hat{X}^{(1)}_{t+\tau}, \ldots, \hat{X}^{(N)}_{t+\tau}\}$ and learns horizon-specific weights, so the resulting ensemble is specialised for each lead time.

\paragraph{FTAE-Cascade$^{Unify}_{\tau}$ and FTAE-Cascade$^{Refine}_{\tau}$.}
As an alternative, we first train a 6-hour base ensemble (FTAE-Unify$_{6}$ or FTAE-Refine$_{6}$) and freeze its weights.
Extended forecasts are then produced by iteratively applying the frozen 6-hour ensemble: the output $\hat{X}_{t+6k}$ at each step feeds back as initial conditions for the next, with $k=1,2,\ldots,\tau/6$.
This cascade preserves temporal consistency and keeps the short-horizon weight patterns interpretable, at the cost of forgoing horizon-specific weight adaptation at longer ranges.

\subsection*{Strategic evolution via Evolve-Agent}
The Evolve-Agent updates $\mathcal{M}_{\mathrm{active}}$ on a slower timescale to keep the model pool current.
Its signal comes directly from the Weight-Agent: models that consistently receive high weights demonstrate practical value, while those assigned low weights contribute little.
After each evaluation epoch, the Evolve-Agent ranks models by their fitness score $\mathcal{F}_i$ (Eq.~\ref{eq:fitness_s1}) and applies a prune-and-inject protocol that removes the lowest-ranked models and replaces them with candidates from the reserve pool $\mathcal{M} \setminus \mathcal{M}_{\mathrm{active}}$.
This feedback loop ties tactical weight allocation to strategic model curation, allowing the system to retire outdated models and incorporate newly released architectures without manual intervention.

\clearpage
\section*{Results}
\label{sec:results}

A comprehensive evaluation of RMSE indicates that the predictive performance of different models varies substantially with forecast lead time. At short lead times, a single strong baseline model remains competitive; however, as the lead time extends, fusion‑based methods, particularly FTAE‑Cascade, gradually demonstrate superior performance, with notably lower prediction errors.

\subsection*{\textbf{Short-Range Forecast Evaluation}
}
\label{subsec:baseline}

\begin{figure}[htbp]
\centering
\includegraphics[width=\linewidth]{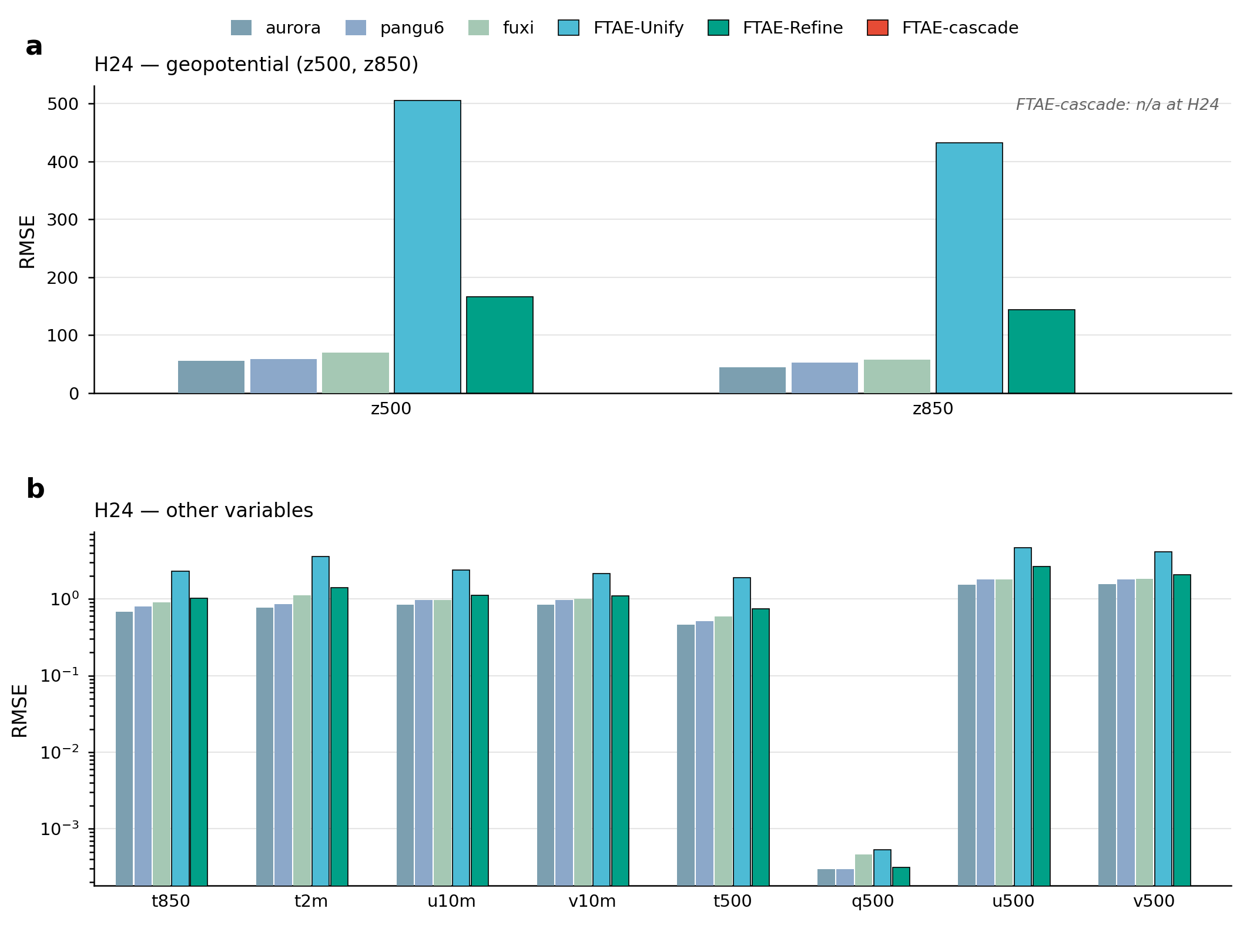}
\caption{Model comparison at a lead time of 24 h. The figure shows the latitude-weighted RMSEs of different models at a lead time of 24 h. Panel a presents z500 and z850 geopotential height variables on a linear scale, while panel b presents t850, t2m, u10m, v10m, t500, q500, u500, and v500 on a logarithmic scale. Bars with different colors denote different models, as indicated in the legend.}
\label{fig:fig1}
\end{figure}

At the 24 h lead time, Aurora achieves the lowest RMSE across all evaluated variables (Fig.~\ref{fig:fig1}). For geopotential height, its errors for z500 and z850 are clearly lower than those of Pangu6, FuXi, and the two direct FTAE variants. Similar behavior is observed for temperature, wind, and humidity variables. In contrast, FTAE-Unify exhibits relatively large errors at this short lead time, especially for z500, z850, and wind-related variables. FTAE-Refine reduces the error relative to FTAE-Unify but still does not outperform the best individual model. This indicates that, at short forecast horizons, model errors have not yet diverged sufficiently for fusion to provide clear additional benefits.

\begin{figure}[htbp]
\centering
\includegraphics[width=\linewidth]{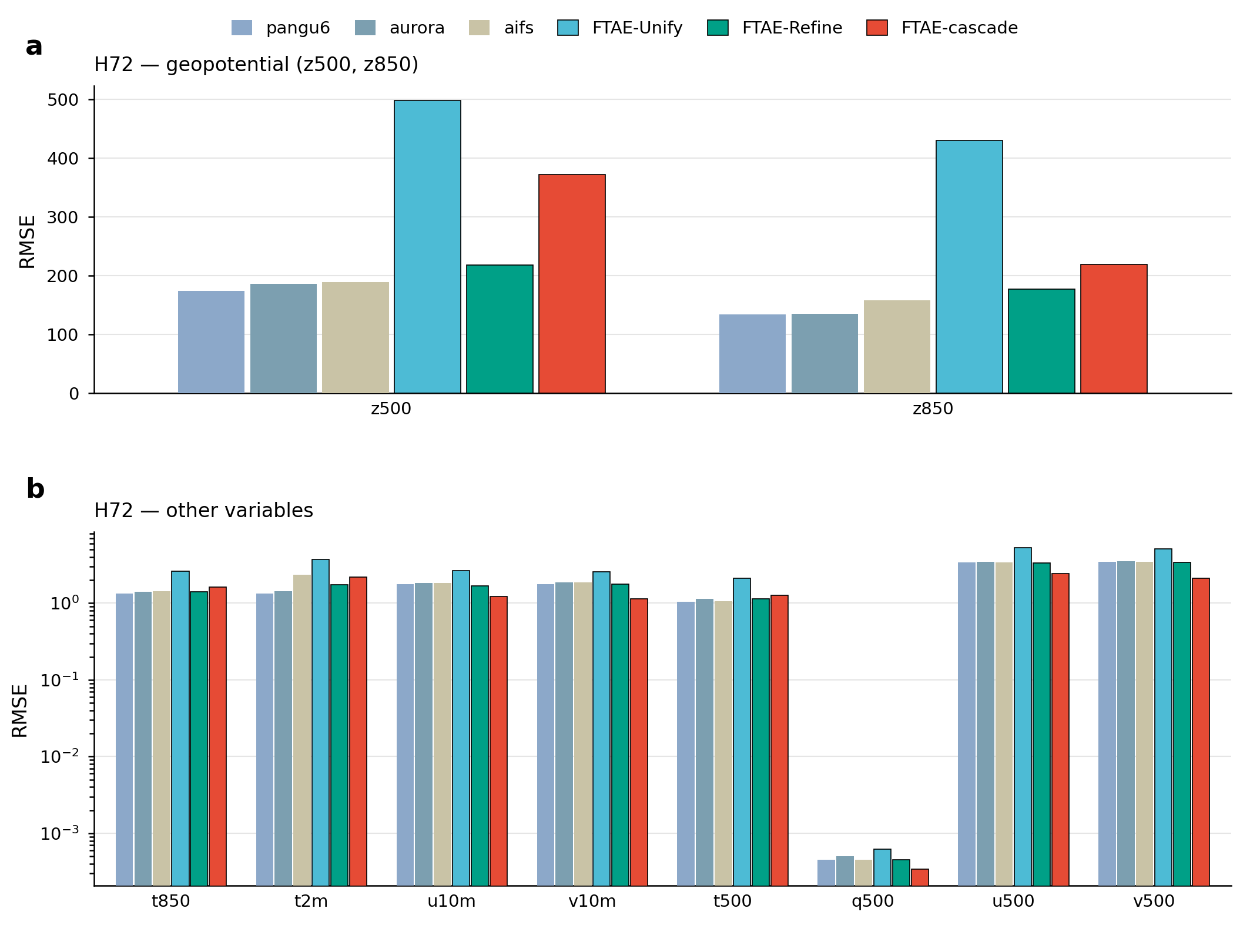}
\caption{Model comparison at a lead time of 72 h. The figure shows the latitude-weighted RMSEs of different models at a lead time of 72 h. Panel a presents z500 and z850 geopotential height variables on a linear scale, while panel b presents t850, t2m, u10m, v10m, t500, q500, u500, and v500 on a logarithmic scale. Bars with different colors denote different models, as indicated in the legend.}
\label{fig:fig2}
\end{figure}

\subsection*{\textbf{Intermediate-Range Forecast Evaluation}
}
\label{subsec:baseline}

At the 72 h lead time, the model ranking becomes more variable-dependent (Fig.~\ref{fig:fig2}). Pangu6 performs well for geopotential height and several thermodynamic variables, whereas FTAE-Cascade begins to achieve lower errors for wind and humidity variables, including u10m, v10m, q500, u500, and v500. FTAE-Refine also substantially reduces the errors of FTAE-Unify and becomes competitive with the baseline models for some variables. However, its advantage is not yet consistent across all variables. These results suggest that model complementarity starts to emerge at this lead time, but the best individual model remains highly competitive for geopotential height and temperature-related variables.

\begin{figure}[htbp]
\centering
\includegraphics[width=\linewidth]{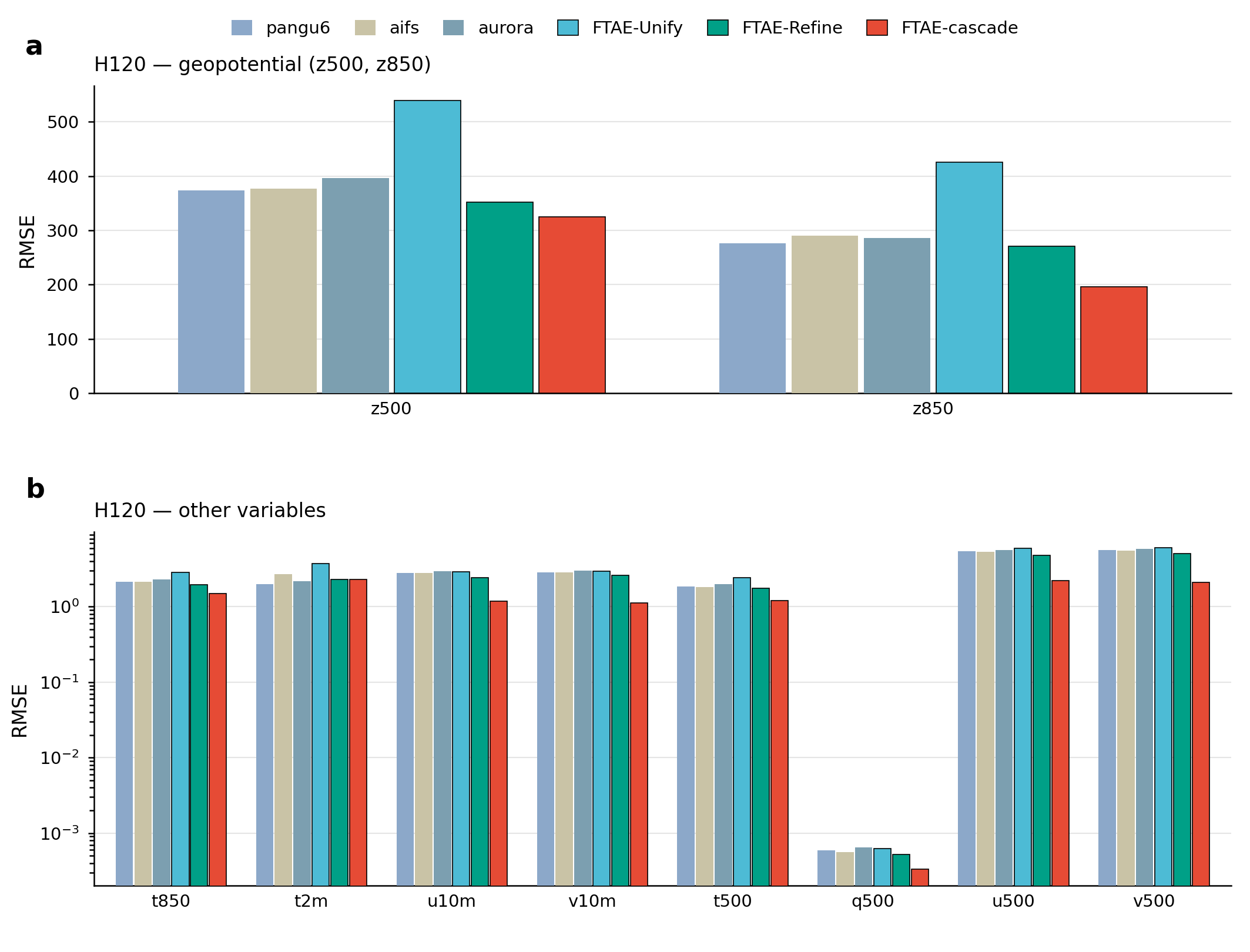}
\caption{Model comparison at a lead time of 120 h. The figure shows the latitude-weighted RMSEs of different models at a lead time of 120 h. Panel a presents z500 and z850 geopotential height variables on a linear scale, while panel b presents t850, t2m, u10m, v10m, t500, q500, u500, and v500 on a logarithmic scale. Bars with different colors denote different models, as indicated in the legend.}
\label{fig:fig3}
\end{figure}

At the 120 h lead time, the benefit of fusion becomes more evident (Fig.~\ref{fig:fig3}). FTAE-Refine improves upon the best baseline model for most variables, indicating that the refined fusion strategy can exploit complementary information among different models. FTAE-Cascade further strengthens this advantage and achieves the lowest RMSE for 9 of the 10 evaluated variables, with t2m being the only exception. Its improvement is particularly clear for geopotential height, wind, and humidity variables. This suggests that, at approximately five days of forecast lead time, the advantage of any single model becomes less stable, while cascade fusion can more effectively integrate variable-dependent model strengths.

\begin{figure}[htbp]
\centering
\includegraphics[width=\linewidth]{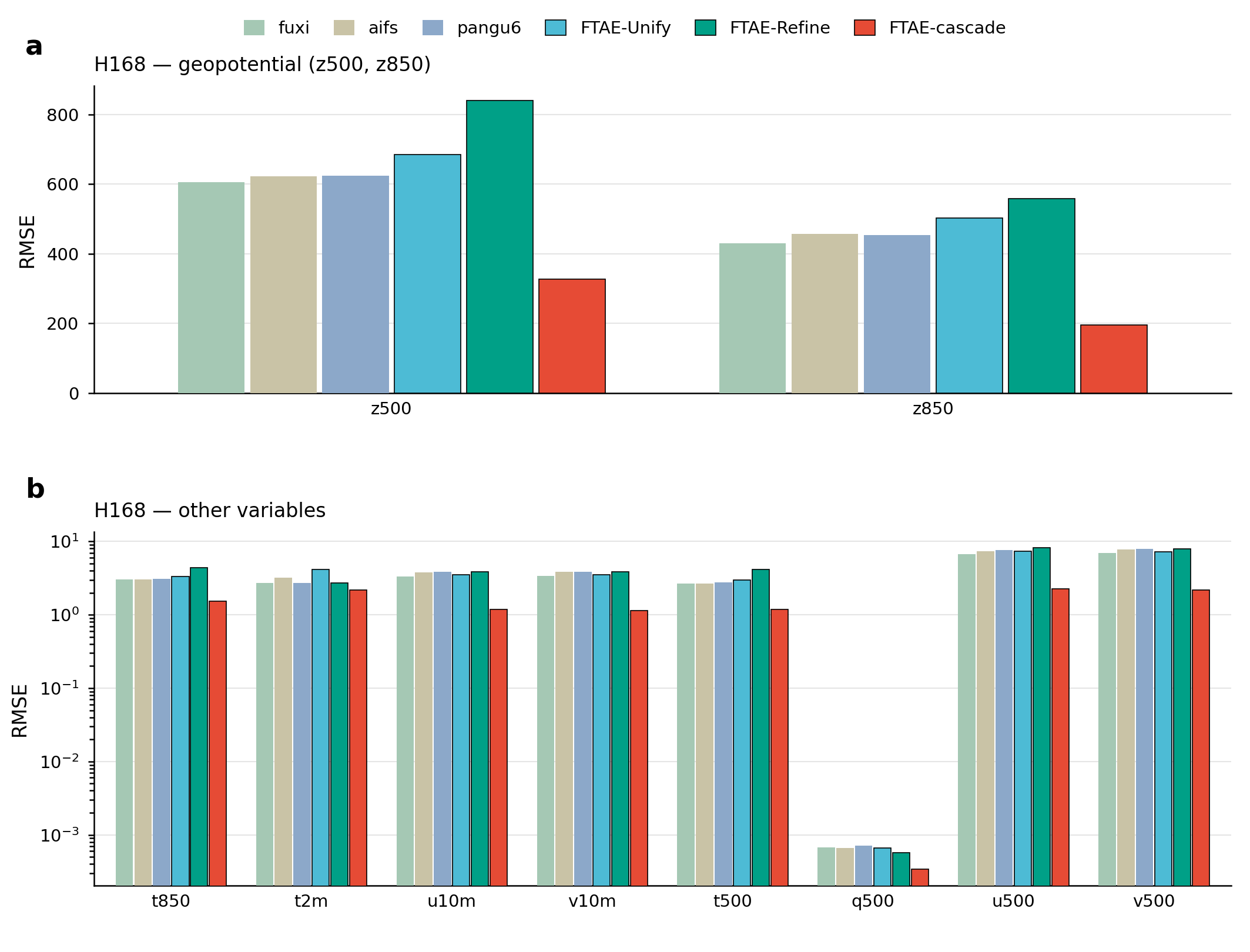}
\caption{\textbf{\textbf{Model comparison at a lead time of 168 h.}} The figure shows the latitude-weighted RMSEs of different models at a lead time of 168 h. Panel a presents z500 and z850 geopotential height variables on a linear scale, while panel b presents t850, t2m, u10m, v10m, t500, q500, u500, and v500 on a logarithmic scale. Bars with different colors denote different models, as indicated in the legend.}
\label{fig:fig4}
\end{figure}

At the 168 h lead time, the advantage of cascade-based fusion becomes more pronounced (Fig.~\ref{fig:fig4}). Among the baseline models, FuXi performs relatively well for most variables, while AIFS and Pangu6 remain competitive for selected variables. In contrast, FTAE-Unify and FTAE-Refine do not consistently outperform the best baseline model at this lead time. In particular, FTAE-Refine shows notably larger errors for geopotential height, suggesting that direct fusion or one-step refined fusion may still be affected by systematic model biases and accumulated forecast errors at 168 h. By contrast, FTAE-Cascade achieves the lowest RMSE across all evaluated variables. Its errors for z500 and z850 are substantially lower than those of both the baseline models and the direct fusion variants, and it also shows clear advantages for temperature, wind, and humidity variables. These results indicate that, at the 7-day forecast range, model error structures have become sufficiently differentiated, allowing cascade fusion to more effectively exploit inter-model complementarity and suppress error growth in both dynamical and thermodynamic fields.

\begin{figure}[htbp]
\centering
\includegraphics[width=\linewidth]{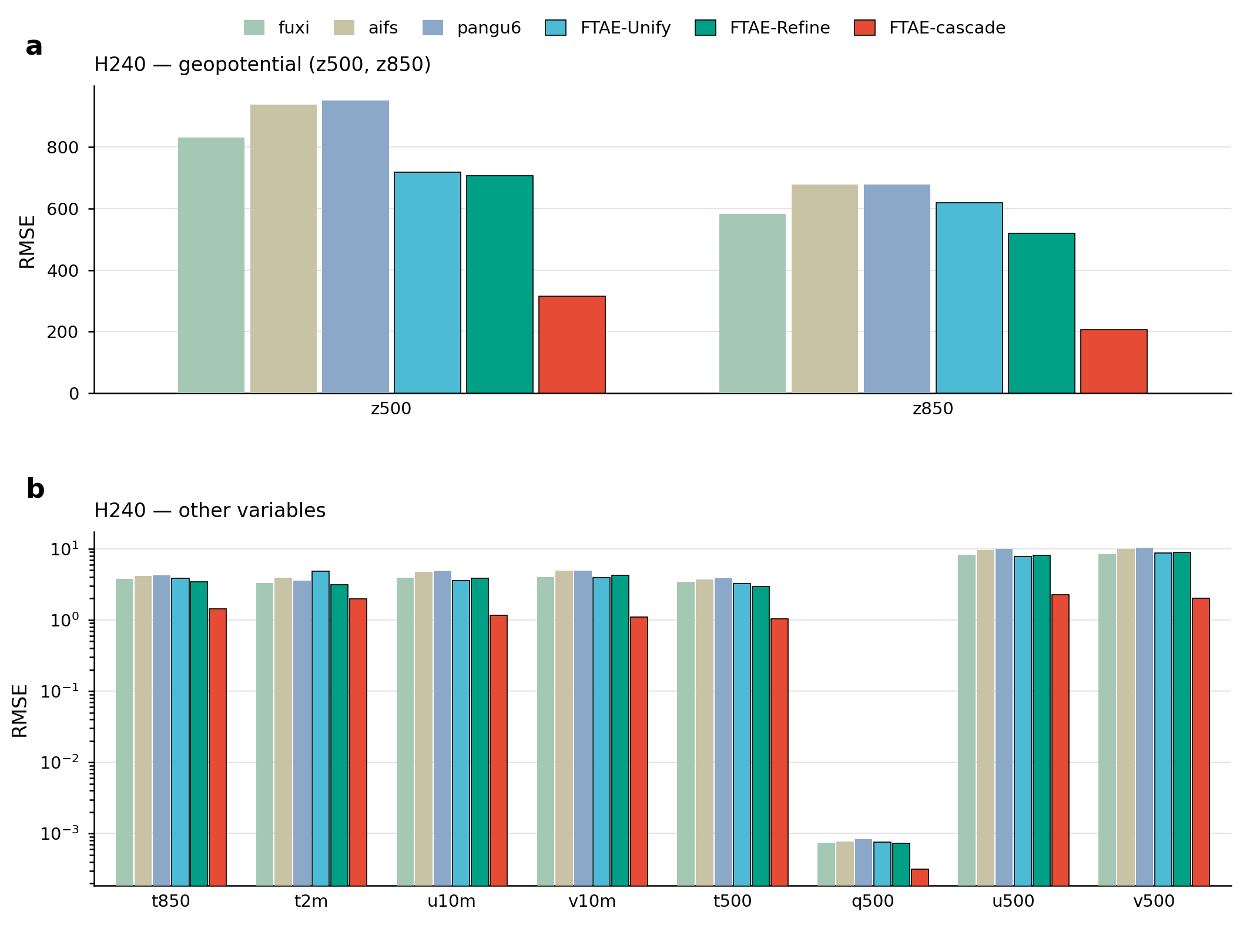}
\caption{\textbf{\textbf{Model comparison at a lead time of 240 h.}} The figure shows the latitude-weighted RMSEs of different models at a lead time of 240 h. Panel a presents z500 and z850 geopotential height variables on a linear scale, while panel b presents t850, t2m, u10m, v10m, t500, q500, u500, and v500 on a logarithmic scale. Bars with different colors denote different models, as indicated in the legend.}
\label{fig:fig5}
\end{figure}

\subsection*{\textbf{Long-Range Forecast Evaluation}
}

At the 240 h lead time, errors in the baseline models increase substantially, especially for geopotential height and upper-level wind variables (Fig.~\ref{fig:fig5}). Among the baseline models, FuXi remains relatively competitive for several variables, but it is consistently outperformed by FTAE-Cascade. Cascade achieves the lowest RMSE across all 10 evaluated variables, including z500, z850, temperature, humidity, and wind variables. FTAE-Refine also improves upon the best baseline model for most variables, although its error reduction remains smaller than that of Cascade. These results indicate that cascade-based fusion provides stronger control of accumulated forecast errors at long lead times.

\begin{figure*}[htbp]
\centering
\includegraphics[width=\linewidth]{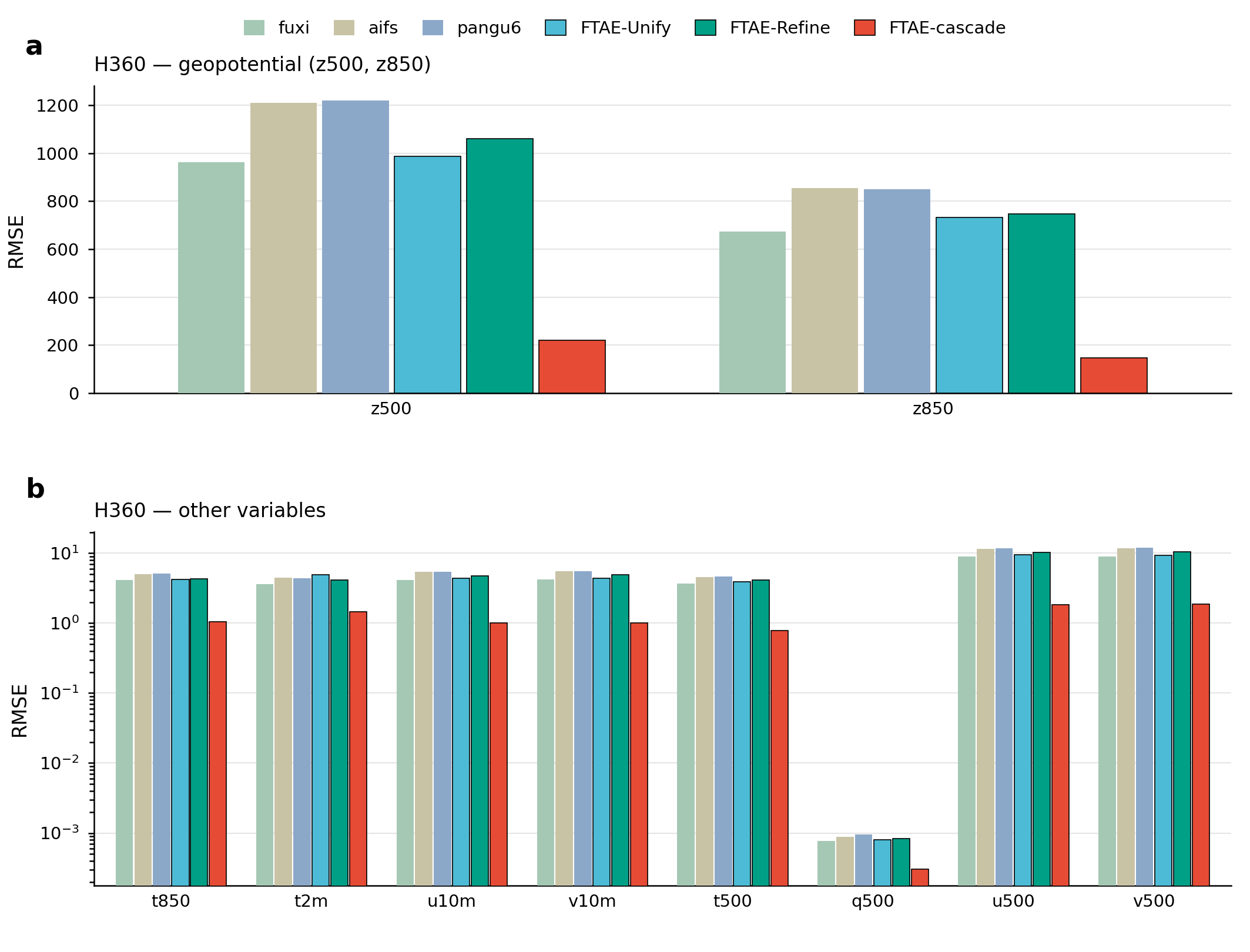}
\caption{Model comparison at a lead time of 360 h. The figure shows the latitude-weighted RMSEs of different models at a lead time of 360 h. Panel a presents z500 and z850 geopotential height variables on a linear scale, while panel b presents t850, t2m, u10m, v10m, t500, q500, u500, and v500 on a logarithmic scale. Bars with different colors denote different models, as indicated in the legend.}
\label{fig:fig6}
\end{figure*}

At the 360 h lead time, the performance gap between FTAE-Cascade and the other methods becomes more pronounced (Fig.~\ref{fig:fig6}). Although FuXi remains one of the stronger baseline models, it does not achieve the lowest RMSE for any evaluated variable. FTAE-Cascade maintains the lowest errors for geopotential height, temperature, humidity, and wind variables. By comparison, FTAE-Unify and FTAE-Refine do not consistently outperform the best baseline model at this lead time, suggesting that direct fusion alone may still be affected by accumulated error propagation and systematic model biases in ultra-long-range forecasts.

\subsection*{\textbf{Lead-Time-Dependent Benefits of Multi-Model Fusion}
}
\label{subsec:baseline}

Across different forecast lead times, the evaluation results reveal a clear lead-time dependence in the benefits of multi-model fusion. At the short lead time of H24, the best individual model, Aurora, remains dominant, indicating that a single high-performing model still benefits from strong initial-condition constraints and short-range stability. By H72, inter-model complementarity begins to emerge: FTAE-Cascade shows advantages for wind and humidity variables, while lower errors in geopotential height and temperature variables are still mainly achieved by the best baseline models. After H120, the benefits of fusion become more evident, with FTAE-Cascade outperforming the baseline models for most variables. This suggests that, as the forecast horizon extends, the error structures of different models become increasingly differentiated, allowing fusion strategies to more effectively exploit complementary information across models.

At the H168 lead time, the advantage of cascade fusion further increases. FTAE-Cascade achieves the lowest RMSE across all evaluated variables, with particularly clear error reduction for geopotential height, wind, and humidity variables. When the lead time is further extended to H240 and H360, the errors of the baseline models continue to increase, whereas FTAE-Cascade maintains the lowest RMSE for all variables, indicating more stable error control in medium- to long-range forecasting.

These results indicate that multi-model fusion does not necessarily outperform the best individual model at all forecast lead times. Its primary advantage is concentrated in medium- to long-range prediction. As the forecast horizon increases, different models exhibit increasingly distinct error-growth characteristics and systematic biases across variables. The cascade fusion strategy can better exploit this complementarity, thereby suppressing accumulated errors in multi-variable forecasts and improving the overall stability of long-range prediction.

\section*{Discussion}
\label{section:discussion}

Our results show that the growing diversity of AI weather models is not a problem to be solved by picking a winner; it is a resource to be exploited.
By learning to read the atmosphere and allocate trust accordingly, FTAE-Weather extracts consistent gains over both individual forecasters and conventional ensemble techniques across variables and lead times spanning 24 to 360~hours.
Three aspects of the framework merit discussion.

\paragraph{Why adaptive weighting outperforms static combination.}
Traditional ensemble methods assign fixed or slowly updated weights that cannot respond to the atmospheric state at forecast time\cite{hagedorn2005rationale,tebaldi2007multimodel,weigel2010risks}.
FTAE-Weather generates weights conditioned on the current analysis and the full set of model forecasts, so it can, for example, favour one architecture during tropical convection and another during midlatitude baroclinic development.
The widening performance gap between FTAE-Refine and simple averaging at longer lead times suggests that regime-dependent model selection becomes increasingly important as forecast uncertainty grows and model predictions diverge\cite{buizza2005comparison}.
This state-dependent gating is conceptually related to the mixture-of-experts principle\cite{jacobs1991adaptive,shazeer2017outrageously}, but differs in that the gating mechanism is learned via temporal-difference reinforcement learning\cite{sutton2018reinforcement} rather than supervised gradient descent, and the experts are frozen pretrained models rather than jointly trained sub-networks.

\paragraph{Limitations and scope.}
The framework's skill ceiling is bounded by the collective capability of the model pool: if all constituent models share a systematic bias (for instance, underestimating polar warming or failing to resolve tropical cyclone intensification), the ensemble inherits that bias, and the residual correction module can only partially compensate.
Our evaluation uses deterministic metrics (RMSE, ACC).
Extending FTAE-Weather to probabilistic verification\cite{wilks2011statistical} (CRPS, reliability diagrams) and to extreme-event prediction (tropical cyclone intensity, heat waves) would directly address the assessment needs of operational forecast centres.
The diffusion-based member of the pool (GenCast\cite{price2025gencast}) already provides probabilistic output that could seed a calibrated ensemble spread, but fully exploiting distributional information within the fusion step would require moving beyond point-estimate weighting.
The Evolve-Agent currently ranks models by a single fitness score; diversity-aware selection\cite{yuksel2012twenty} or multi-objective optimisation balancing skill against computational cost could improve pruning decisions.
Finally, while we train and evaluate on ERA5 reanalysis, operational deployment would require coupling with real-time initial conditions and accounting for observation-system changes that may shift the relative strengths of constituent models over time.

\paragraph{Relation to operational practice.}
Operational centres already run multi-model ensembles\cite{buizza2005comparison,hagedorn2005rationale}, but their combination rules are typically fixed over a forecast season.
FTAE-Weather's per-cycle weight adaptation could complement existing practice by providing a data-driven second opinion that highlights when the standard combination deviates from an optimal weighting for the prevailing regime.
The lightweight parameter footprint and asynchronous caching make it feasible to run alongside operational pipelines without significant additional computational burden.
However, integrating a learned fusion layer into safety-critical workflows would require thorough reliability assessment and clear protocols for handling edge cases where the agent assigns extreme weights to a single model.

\paragraph{Broader implications.}
Architectural diversity in AI weather forecasting need not be a coordination problem.
Every new specialist model, regardless of design philosophy or training institution, can be absorbed into an FTAE ensemble and improve the collective forecast at negligible parameter cost ($<$0.01\% of a typical foundation model).
As data-driven weather forecasting continues to diversify across architectures\cite{nguyen2023climax,keisler2022forecasting}, data sources, and target applications\cite{ben2024rise,charlton2024ai}, coordination frameworks of this kind may prove more sustainable than the pursuit of ever-larger monolithic models.
FTAE-Weather offers a concrete realisation of this principle.

\clearpage
\section*{Methods}
\label{sec:methods}

\subsection*{Forecasting task and notation}
\label{subsec:method_notation}

The global atmospheric state at time $t$ is represented as $X_t \in \mathbb{R}^{H \times W \times V}$ on a $0.25^{\circ}$ latitude--longitude grid ($H{=}721$, $W{=}1440$, $V{=}69$ variables).
Given initialisation time $t_0$, the goal is to predict $X_{t_0+\tau}$ at lead time $\tau \in \{6, 12, 24, 72, 120, 168, 240, 360\}$~h.
We assume access to a pool of pretrained forecasters $\mathcal{M}=\{M^{(1)},\dots,M^{(N_{\mathrm{pool}})}\}$, each mapping $\mathbb{R}^{H \times W \times V} \rightarrow \mathbb{R}^{H \times W \times V}$.
An active subset $\mathcal{M}_{\mathrm{active}}\subseteq\mathcal{M}$ with $N = |\mathcal{M}_{\mathrm{active}}|$ models participates at any given time.
The fusion is parameterised by $\Phi=\{W^{(1)},\dots,W^{(N)},b\}$, where $W^{(i)} \in \mathbb{R}^{H \times W \times V}$ is the weight field for model $i$ and $b \in \mathbb{R}^{H \times W \times V}$ is an additive bias correction.
The fused forecast is
\begin{equation}
    \hat{X}_{t_0+\tau}
    =
    \sum_{i=1}^{N}
    W^{(i)}_{t_0} \odot M^{(i)}(X_{t_0})
    +
    b_{t_0},
    \label{eq:tactical_fusion_methods}
\end{equation}
where $\odot$ denotes element-wise multiplication, $W^{(i)}_{t_0}(h,w,v) \ge 0$ for all spatial locations $(h,w)$ and variables $v$, with the simplex constraint $\sum_{i=1}^{N} W^{(i)}_{t_0}(h,w,v) = 1$ enforced at each grid point and variable.

Optimisation proceeds on two timescales.
At the \emph{tactical} level, weight generation is cast as an MDP in which a policy network $\pi_{\phi}$ produces $\Phi$ conditioned on the atmospheric state and receives verification-based rewards.
At the \emph{strategic} level, $\mathcal{M}_{\mathrm{active}}$ is updated periodically.
Model utility is measured by the tactical controller's \emph{revealed preference}, that is, the accumulated weight allocation over an evaluation window $T_{\mathrm{eval}}$, which defines a fitness score
\begin{equation}
\mathcal{F}_i
=
\frac{1}{T_{\mathrm{eval}}}
\sum_{t=1}^{T_{\mathrm{eval}}}
\left\langle W^{(i)}_t \right\rangle,
\label{eq:fitness_s1}
\end{equation}
where $\langle W^{(i)}_t\rangle = \frac{1}{V}\sum_{v=1}^{V} \bar{W}^{(i)}_t(v)$ denotes the variable-averaged weight assigned to expert $i$ at time $t$.
%
This fitness metric drives a prune-and-inject protocol that updates the active expert set.

\subsection*{Data preparation}
\label{subsec:data}

All experiments use the ERA5 reanalysis\cite{hersbach2020era5} at $0.25^{\circ}$ resolution.
Upper-air variables (geopotential height, temperature $T$, specific humidity $q$, wind components $u$, $v$) are included at 13 pressure levels ($p \in \{50, 100, 150, 200, 250, 300, 400, 500, 600, 700, 850, 925, 1000\}$~hPa), alongside surface fields ($T_{2m}$, $u_{10}$, $v_{10}$, mean sea-level pressure).
Data are split chronologically: 1980--2017 for training, 2018--2019 for validation, and 2020--present for testing.
ACC is computed against a daily climatology over 1991--2020, smoothed with a 7-day centred moving average.
All input fields are standardised to zero mean and unit variance using training-period statistics only.

\subsection*{Evaluation metrics}
\label{subsec:metrics}

Forecast skill is measured by latitude-weighted RMSE and ACC\cite{wilks2011statistical,murphy1993good}.
For forecast $\hat{Y} \in \mathbb{R}^{H \times W}$ and reference $Y \in \mathbb{R}^{H \times W}$, the latitude-weighted RMSE is
\begin{equation}
    \mathrm{RMSE} = \sqrt{\frac{\sum_{h=1}^{H}\sum_{w=1}^{W} (\hat{Y}_{h,w} - Y_{h,w})^2 \cos(\phi_h)}{\sum_{h=1}^{H}\sum_{w=1}^{W} \cos(\phi_h)}},
\end{equation}
where $\phi_h$ denotes the latitude at grid row $h$.
ACC is the latitude-weighted Pearson correlation between forecast anomalies $\hat{Y}' = \hat{Y} - \bar{Y}$ and observed anomalies $Y' = Y - \bar{Y}$ (with $\bar{Y}$ the climatological mean):
\begin{equation}
    \mathrm{ACC} = \frac{\sum_{h,w} (\hat{Y}'_{h,w} \cdot Y'_{h,w}) \cos(\phi_h)}{\sqrt{\sum_{h,w} (\hat{Y}'_{h,w})^2 \cos(\phi_h)} \sqrt{\sum_{h,w} (Y'_{h,w})^2 \cos(\phi_h)}}.
\end{equation}

\subsection*{Expert model zoo}
\label{subsec:expert_models}

Each model in $\mathcal{M}$ is accessed as a black-box forecaster through a common input--output interface.
In this study, the pool consists of eight pretrained models: GenCast, GraphCast, Pangu-Weather, Aurora, FuXi, AIFS, FourCastNet3, and SFNO.
These models cover several representative families of AI weather forecasters, including Transformer-based models (Pangu-Weather\cite{bi2023panguweather}, Aurora\cite{bodnar2024aurora}, FuXi\cite{chen2023fuxi}), graph-based models (GraphCast\cite{lam2023graphcast}, AIFS\cite{lang2024aifs}), diffusion models (GenCast\cite{price2025gencast}), and neural-operator or spectral-operator models (FourCastNet3\cite{bonev2025fourcastnet3}, SFNO\cite{bonev2023sfno}).
All outputs are interpolated to the common ERA5 grid ($0.25^{\circ}$) and aligned to a shared variable set, pressure-level set, lead-time convention, and unit system before fusion.
For probabilistic models such as GenCast, deterministic summaries such as the ensemble mean are used when RMSE- or ACC-based fusion and evaluation are performed.
Predictions are cached asynchronously during policy training, so that Weight-Agent updates are decoupled from repeated execution of the slowest constituent model.

\subsection*{Weight-Agent}
\label{subsec:weight_agent}

The Weight-Agent is an off-policy continuous-control agent that predicts dense weight fields $W^{(i)} \in \mathbb{R}^{H \times W \times V}$ rather than scalar mixture coefficients.
A lightweight CNN encoder maps the high-resolution atmospheric input ($721 \times 1440$ grid points) to a compact latent representation, and a deterministic actor--critic architecture\cite{lillicrap2015continuous,fujimoto2018addressing} with experience replay drives learning.
Table~\ref{tab:arch_specs} lists the main components.

\begin{table}[t]
\centering
\captionsetup{labelformat=empty,textformat=simple}
\caption{\textbf{Network architecture specifications used in FTAE-Weather.}}
\begin{tabular}{@{}llll@{}}
\toprule
\textbf{Component} & \textbf{Module} & \textbf{Specification} & \textbf{Activation} \\
\midrule
State encoder & CNN feature extractor & Stacked Conv/SepConv blocks; multi-scale features & ReLU  \\
(Backbone) & Global pooling & AdaptiveAvgPool $\rightarrow$ latent $h_{t_0} \in \mathbb{R}^{d}$ & Linear  \\
\midrule
Actor head & Lightweight decoder & Output $H{\times}W{\times}V{\times}N$ logits & Softmax over $N$ \\
\midrule
Critic & MLP & Input: pooled $h_{t_0}$ + compact action embedding & ReLU / Linear \\
\midrule
Training & Replay buffer & Store $(s,a,r,s')$; uniform mini-batch sampling & -- \\
\bottomrule
\multicolumn{4}{p{13cm}}{\footnotesize{\textit{Note:} $N = |\mathcal{M}_{\mathrm{active}}|$ is the number of active models; $(H,W,V) = (721, 1440, 69)$ denote grid dimensions and variable count.
The CNN encoder maintains computational tractability; the decoder restores full-resolution weight maps.}}
\end{tabular}
\label{tab:arch_specs}
\end{table}

\paragraph{Markov decision process.}
Weight generation is formalised as an MDP $\langle\mathcal{S},\mathcal{A},\mathcal{P},r,\gamma\rangle$.
The state $s_{t_0} \in \mathcal{S}$ concatenates the initial conditions $X_{t_0}$ and expert predictions $\{M^{(i)}(X_{t_0})\}_{i=1}^{N}$.
The action $a_{t_0}=\big(W^{(1)}_{t_0},\ldots,W^{(N)}_{t_0}\big)$ satisfies $W^{(i)}_{t_0}(h,w,v)\ge 0$ and $\sum_{i} W^{(i)}_{t_0}(h,w,v)=1$ everywhere.
Transitions are deterministic, advancing to the next initialisation cycle.

\paragraph{Reward design.}
Once verification data at $t_0+\tau$ are available, a composite reward combining three terms is computed:
\begin{equation}
r_{t_0+\tau}
=
\alpha\,r^{\mathrm{skill}}_{t_0+\tau}
+
\beta\,r^{\mathrm{rank}}_{t_0+\tau}
+
\gamma_r\,r^{\mathrm{gain}}_{t_0+\tau},
\label{eq:method_reward}
\end{equation}
with $\alpha=1.0$, $\beta=0.5$, $\gamma_r=0.8$ (set on the validation period; note that $\gamma_r$ denotes the gain-term coefficient to distinguish it from the MDP discount factor $\gamma$).
All terms are oriented so that higher values indicate better forecasts.
The skill term uses the quantile-transformed symmetric mean absolute percentage error (sMAPE):
\begin{equation}
r^{\mathrm{skill}}_{t_0+\tau} = 1 - \Phi_{\mathrm{sMAPE}}\left(\frac{1}{HWV}\sum_{h,w,v}\frac{2|\hat{X}_{t_0+\tau}(h,w,v) - X_{t_0+\tau}(h,w,v)|}{|\hat{X}_{t_0+\tau}(h,w,v)| + |X_{t_0+\tau}(h,w,v)|}\right),
\label{eq:reward_skill}
\end{equation}
where $\Phi_{\mathrm{sMAPE}}$ is the CDF fitted on training-period sMAPE values; lower error maps to higher reward.
The rank term scores the ensemble's ordinal position among the $N$ experts plus itself:
\begin{equation}
r^{\mathrm{rank}}_{t_0+\tau} = 1 - \frac{\mathrm{rank}(\hat{X}_{t_0+\tau})}{N+1},
\label{eq:reward_rank}
\end{equation}
where rank 1 is best (lowest RMSE).
The gain term rewards improvement over the best individual model:
\begin{equation}
r^{\mathrm{gain}}_{t_0+\tau} = \max\left(0, \frac{\min_{i}\mathrm{RMSE}(M^{(i)}(X_{t_0}), X_{t_0+\tau}) - \mathrm{RMSE}(\hat{X}_{t_0+\tau}, X_{t_0+\tau})}{\min_{i}\mathrm{RMSE}(M^{(i)}(X_{t_0}), X_{t_0+\tau})}\right),
\label{eq:reward_gain}
\end{equation}
This term is positive when the ensemble beats every expert and is clipped to zero otherwise.

\paragraph{State encoding.}
A lightweight CNN backbone\cite{KrizhevskyCNN2012} extracts multiscale features from $X_{t_0}$, which are aggregated by adaptive global average pooling into a latent vector $h_{t_0} \in \mathbb{R}^{d}$ ($d{=}512$) shared by the actor and critic.

\paragraph{Policy optimisation.}
A replay buffer $\mathcal{B}$ stores $(s_{t_0},a_{t_0},r_{t_0},s'_{t_0})$ tuples\cite{mnih2015humanlevel}.
During collection, the actor adds Gaussian exploration noise: $a_{t_0}=\pi_{\phi}(s_{t_0})+\epsilon_{t_0}$, $\epsilon_{t_0} \sim \mathcal{N}(0,\sigma^2)$ with $\sigma=0.1$.
Mini-batches sampled uniformly from $\mathcal{B}$ break temporal correlations\cite{schulman2017proximal}.
Target networks are updated via Polyak averaging ($\rho{=}0.005$): $\theta_{\mathrm{target}} \leftarrow \rho\,\theta + (1{-}\rho)\,\theta_{\mathrm{target}}$.

\paragraph{Bias correction.}
For FTAE-Refine, the weighted fusion $\tilde{X}_{t_0+\tau}=\sum_{i} W^{(i)}_{t_0} \odot M^{(i)}(X_{t_0})$ is corrected by an additive residual $b_{t_0}$ predicted from $h_{t_0}$ by a small MLP trained with $\mathcal{L}_{\mathrm{bias}} = \|b_{t_0} - b^{\star}_{t_0}\|^2_2$, where $b^{\star}_{t_0}=X_{t_0+\tau}-\tilde{X}_{t_0+\tau}$.
The final forecast is $\hat{X}_{t_0+\tau}=\tilde{X}_{t_0+\tau}+b_{t_0}$.
For FTAE-Unify, $b_{t_0} = 0$ because uniform weights across variables make per-variable bias correction ill-defined.

\subsection*{Evolve-Agent}
\label{subsec:evolve_agent}

The Evolve-Agent manages $\mathcal{M}_{\mathrm{active}}$ on a slower timescale, using the Weight-Agent's revealed preferences rather than a separate neural policy.

\paragraph{Fitness evaluation.}
Over an evaluation epoch of duration $T_{\mathrm{eval}}$ (typically several months), the variable-averaged weight for expert $i$ at time $t$ is
\begin{equation}
\bar{W}^{(i)}_{t} \;=\; \frac{1}{V}\sum_{v=1}^{V} \left(\frac{1}{HW}\sum_{h=1}^{H}\sum_{w=1}^{W} W^{(i)}_{t}(h,w,v)\right),
\end{equation}
and the fitness score is the temporal average:
\begin{equation}
\mathcal{F}_i \;=\; \frac{1}{T_{\mathrm{eval}}}\sum_{t=1}^{T_{\mathrm{eval}}}\bar{W}^{(i)}_{t}.
\label{eq:fitness}
\end{equation}
High $\mathcal{F}_i$ indicates consistent utilisation; low $\mathcal{F}_i$ signals marginal contribution.

\paragraph{Prune-and-inject protocol.}
After each epoch, the bottom $k$ models (typically $k{=}2$--$3$) are removed and replaced by candidates from $\mathcal{M}\setminus\mathcal{M}_{\mathrm{active}}$, prioritising newly released architectures or updated checkpoints.
The Weight-Agent continues without interruption, reallocating weights as the pool changes.
\clearpage
\section*{Data availability}
All data used in this study are publicly available. ERA5 reanalysis data are provided by ECMWF (\url{https://cds.climate.copernicus.eu}). 
%

\section*{Code availability}
Code for training and evaluating FTAE-Weather will be available.


\bibliography{new}
\section*{Acknowledgments}
This work is supported by National Natural Science Foundation of China (Grant Nos. 42595593) 
, Fundamental and Interdisciplinary Disciplines Breakthrough Plan of the Ministry of Education of China (Grant No. JYB2025XDXM910), and Talent Scientific Fund of Lanzhou University (Grant No. 561120208).

\clearpage
\renewcommand{\thefigure}{S\arabic{figure}}
\renewcommand{\thetable}{S\arabic{table}}
\setcounter{figure}{0}
\setcounter{table}{0}

\end{document}